\documentclass[aps,prl,showpacs,twocolumn,groupedaddress,superscriptaddress,nobibnotes]{revtex4}
\usepackage{amsmath}
\usepackage{amssymb}
\usepackage{graphicx}
\usepackage{dcolumn}
\usepackage{bm}
\usepackage{color}
\usepackage{float}
\usepackage[normalem]{ulem}
\usepackage{empheq}
\usepackage[none]{hyphenat}

\begin{document}

\title{Quasiparticle interference on the surface
      of 3D topological insulator Bi$_{2}$Se$_{3}$ induced by cobalt adatom in the
      absence of ferromagnetic ordering}

\author{M. Ye}
\altaffiliation{Present address: Hiroshima Synchrotron Radiation Center, Hiroshima
University, Japan.}
\affiliation{Graduate School of Science, Hiroshima University, 1-3-1
Kagamiyama, Higashi-Hiroshima, 739-8526, Japan}

\author{S. V. Eremeev}
\affiliation{%
    Institute of Strength Physics and Materials Science, 634021, Tomsk,
    Russia}
\affiliation{%
    Tomsk State University, 634050, Tomsk, Russia}

\author{K. Kuroda}
\affiliation{Graduate School of Science, Hiroshima University, 1-3-1
Kagamiyama, Higashi-Hiroshima, 739-8526, Japan}

\author{E.~E.~Krasovskii}
\affiliation{%
Departamento de F\'{\i}sica de Materiales UPV/EHU and Centro de
F\'{\i}sica de Materiales CFM and Centro Mixto CSIC-UPV/EHU, 20080
San Sebasti\'an/Donostia, Basque Country, Spain
\\
}
\affiliation{%
Donostia International Physics Center (DIPC),
             20018 San Sebasti\'an/Donostia, Basque Country,
             Spain\\
}
\affiliation{%
IKERBASQUE, Basque Foundation for Science, 48011 Bilbao, Spain\\
}

\author{E.~V.~Chulkov}
\affiliation{%
Departamento de F\'{\i}sica de Materiales UPV/EHU and Centro de
F\'{\i}sica de Materiales CFM and Centro Mixto CSIC-UPV/EHU, 20080
San Sebasti\'an/Donostia, Basque Country, Spain
\\
}
\affiliation{%
Donostia International Physics Center (DIPC),
             20018 San Sebasti\'an/Donostia, Basque Country,
             Spain\\
}

\author{Y. Takeda}
\affiliation{Condensed Matter Science Division, Japan Atomic Energy Agency, Sayo, Hyogo 679-5148, Japan}

\author{Y. Saitoh}
\affiliation{Condensed Matter Science Division, Japan Atomic Energy Agency, Sayo, Hyogo 679-5148, Japan}

\author{K. Okamoto}
\affiliation{Graduate School of Science, Hiroshima University, 1-3-1
Kagamiyama, Higashi-Hiroshima, 739-8526, Japan}

\author{S. Y. Zhu}
\affiliation{Graduate School of Science, Hiroshima University, 1-3-1
Kagamiyama, Higashi-Hiroshima, 739-8526, Japan}

\author{K. Miyamoto}
\affiliation{Hiroshima Synchrotron Radiation Center, Hiroshima
University, 2-313 Kagamiyama,  Higashi-Hiroshima, 739-0046, Japan}

\author{M. Arita}
\affiliation{Hiroshima Synchrotron Radiation Center, Hiroshima
University, 2-313 Kagamiyama,  Higashi-Hiroshima, 739-0046, Japan}

\author{M. Nakatake}
\affiliation{Hiroshima Synchrotron Radiation Center, Hiroshima
University, 2-313 Kagamiyama,  Higashi-Hiroshima, 739-0046, Japan}

\author{T. Okuda}
\affiliation{Hiroshima Synchrotron Radiation Center, Hiroshima
University, 2-313 Kagamiyama,  Higashi-Hiroshima, 739-0046, Japan}

\author{Y. Ueda}
\affiliation{Kure National College of Technology, Agaminami 2-2-11,
Kure 737-8506, Japan}

\author{K. Shimada}
\affiliation{Hiroshima Synchrotron Radiation Center, Hiroshima
University, 2-313 Kagamiyama,  Higashi-Hiroshima, 739-0046, Japan}

\author{H. Namatame}
\affiliation{Hiroshima Synchrotron Radiation Center, Hiroshima
University, 2-313 Kagamiyama,  Higashi-Hiroshima, 739-0046, Japan}

\author{M. Taniguchi}
\affiliation{Graduate School of Science, Hiroshima University, 1-3-1
Kagamiyama, Higashi-Hiroshima, 739-8526, Japan}
\affiliation{Hiroshima Synchrotron Radiation Center, Hiroshima
University, 2-313 Kagamiyama,  Higashi-Hiroshima, 739-0046, Japan}

\author{A. Kimura}
\email{akiok@hiroshima-u.ac.jp}
\affiliation{Graduate School of Science, Hiroshima University, 1-3-1 Kagamiyama, Higashi-Hiroshima,
739-8526, Japan}

\date{\today}

\begin{abstract}
Quasiparticle interference induced by cobalt adatoms on the surface of the
topological insulator Bi$_{2}$Se$_{3}$ is studied by scanning tunneling microscopy,
angle-resolved photoemission spectroscopy and X-ray magnetic circular dichroism.
It is found that Co atoms are selectively adsorbed on top of Se sites and act
as strong scatterers at the surface, generating anisotropic standing waves. A
long-range magnetic order is found to be absent, and the surface state Dirac
cone remains gapless. The anisotropy of the standing wave is ascribed to the
heavily warped iso-energy contour of unoccupied states, where the scattering
is allowed due to a non-zero out-of-plane spin.
\end{abstract}

\maketitle
\newpage

A novel class of materials, called topological insulators (TI)
\cite{Hsieh2008nature, Hasan2010RMP, SC.Zhang09NP} with a nontrivial
metallic surface state in the bulk energy gap induced by spin-orbit
coupling have invoked both theoretical and experimental interest..
An odd number of surface states with a spin helical texture established
for Bi$_{1-x}$Sb$_{x}$ \cite{Hasan09_SbBi,Ali09_STM-SbBi}, Bi$_{2}$Te$_{3}$
\cite{ZX.Shen09_BiTe}, Bi$_{2}$Se$_{3}$ \cite{Xia09_BiSe, kuroda2010BiSe},
and Tl-based ternary compounds
\cite{EremeevJETPL_Th_10,Kuroda_PRL10,Eremeev_Tl_PRB2011,Yan_EPL10,Lin_PRL10,Sato_PRL10}
promises a robust protection of spin polarized surface states from
backscattering in the presence of non-magnetic impurities due to
time-reversal (TR) symmetry, which is a key requirement to revolutionize
modern electronic devices. Among discovered TI materials, Bi$_{2}$Se$_{3}$
is one of the most promising candidates owing to the large bulk energy gap
with a single Dirac cone surface state.

In fact, a spin-selective scattering has been directly imaged by scanning
tunneling microscopy (STM) even without the breaking of TR
symmetry~\cite{Ali09_STM-SbBi, Zhang_Xue09_AgBiTe, ZX.shen2010_STM-BiTe}.
Especially for Bi$_{2}$Te$_{3}$, a suppressed backscattering of topological
surface electrons by nonmagnetic impurities has been reported \cite{Zhang_Xue09_AgBiTe}.
It should be noted that owing to the relatively small size of bulk energy
gap and a strong anisotropy of the bulk band structure, the
topological surface state of Bi$_{2}$Te$_{3}$ is strongly hexagonally
warped \cite{ZX.shen2010_STM-BiTe, SC.Zhang09}, which leads to the
quasiparticle interference (QPI) with scattering vector along the
$\bar{\Gamma}$-$\bar{\rm M}$ direction between states with out-of-plane
spin components \cite{Zhang_Xue09_AgBiTe}, even though the backscattering is strictly forbidden due to the TR symmetry. In other words, more favorable
protection would be realized in TI materials with larger
bulk energy gap, such as Bi$_{2}$Se$_{3}$.

Most interesting properties appear when the surfaces of 3D TIs are
interfaced with ferromagnetic layers
\cite{Qi2009Science,Yu2010Science,Biswas2010PRB,Garate2010PRL,MTC11}.
Due to a broken TR symmetry, an energy gap opens at the Dirac point,
leading to massive Dirac fermions. In such a system novel physical
phenomena are expected to emerge, in particular the half quantum Hall effect on
the surface with a Hall conductance of $e^{2}/2h$ \cite{SC.Zhang2008PRB},
as well as the Kondo effect mediated by the Dirac fermions \cite{Wehhling2010PRB}.

Experimentally, great efforts have been made on the magnetic doping in
the bulk of TI materials. An opening of the energy gap at Dirac point and the consequent QPI on the topological surface due to the broken TR symmetry have been directly observed \cite{Chen2010Sci,Okada2011PRL}. On the other hand,
the magnetic doping at the surface, which can be realized by the deposition of magnetic atoms, is still poorly explored although it is expected
to stronger influence the surface Dirac fermions than the bulk doping
\cite{SC.Zhang2009PRL, Wray2010NP}. Here we show for the first time
that Co atoms on the surface of Bi$_{2}$Se$_{3}$ generate a distinct QPI
of the surface Dirac fermions without opening of an energy gap at
the Dirac point. The absence of a long-range ferromagnetic order is
revealed by combined STM, angle-resolved photoemission spectroscopy (ARPES), and x-ray magnetic circular dichroism
(XMCD) experiments.

Our measurements were performed on the pristine Bi$_{2}$Se$_{3}$ single
crystal, which is naturally n-type. Clean surfaces were obtained by
{\it in-situ} cleavage in ultra-high vacuum at room temperature.
Cobalt deposition was made at room temperature after the confirmation
of the clean surface by STM. The present STM results were obtained at
4.7~K in the constant current mode, and the differential conductance
($dI/dV$) map was acquired by a standard lock-in technique. ARPES experiments
were performed at BL-7 of Hiroshima Synchrotron Radiation Center equipped
with a hemispherical photoelectron analyzer (VG-SCIENTA SES2002). XMCD
experiment on Co deposited Bi$_{2}$Se$_{3}$ surface was conducted at the
twin helical undulator beam line BL23SU of SPring-8 in a total-electron yield mode,
employing a superconducting magnet and a helium cryostat.

\begin{figure}
\includegraphics[width=.95\columnwidth]{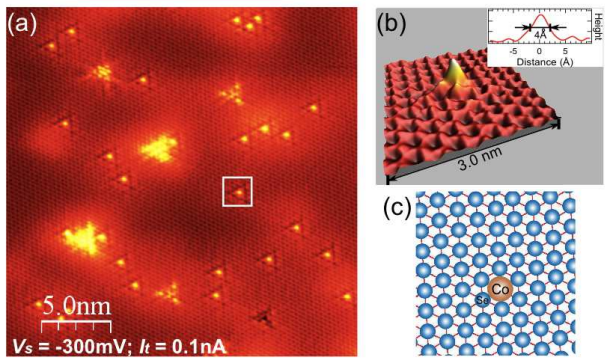}
\caption{\label{fig:epsart0} (Color online) (a) STM image of Co deposited
Bi$_{2}$Se$_{3}$ surface (25nm$\times$25nm). (b) 3D illustration of a
Co adatom (a), inset, cross-sectional profile of the Co adatom on
Bi$_{2}$Se$_{3}$. (c) Schematics of atomic structure in (b).}
\end{figure}

Figure~1(a) shows a typical STM image acquired at a sample bias voltage of
$V_{s}=-300$~mV of the Co-deposited Bi$_{2}$Se$_{3}$ surface. Two kinds of bright
features can be identified in the image, namely the large triangular-shaped
patterns and the tiny bright spots. The large features are typically found in
both Bi$_{2}$Se$_{3}$ and Bi$_{2}$Te$_{3}$ surfaces and are assigned as the
substitutional or antisite defect \cite{diffect2002PRB,Xue2011AdvMat,Kim2011PRL}.
The tiny bright spots appear after the Co deposition and are assigned as Co adatoms.
A close-up STM image of the Co adatom [marked with white frame in Fig.~1(a)] is
shown in Fig.~1(b). The 3D illustration of the Co adatom in Fig.~1(b) clearly
shows the sharp protrusion on the surface with a spatial full width at half
maximum of 4~\AA ~[inset of Fig.~1(b)]. By carefully examining the atomic
position in this area we identify the Co adatom as located on the top of the
Se atom, as schematically shown in Fig.~1(c).

\begin{figure}
\includegraphics[width=\columnwidth]{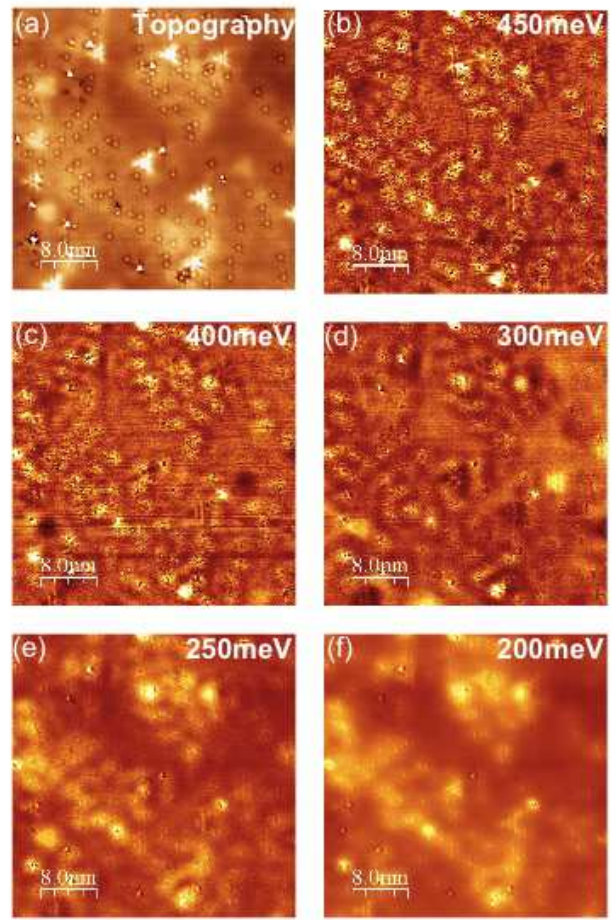}
\caption{\label{fig:epsart1} (Color online) (a) 40nm $\times$ 40nm area
STM image of Co evaporated Bi$_{2}$Se$_{3}$ surface (sample bias
300mV, set point 0.10nA); (b)-(f) $dI/dV$ images taken at the same
area as (a) with different sample bias voltages (set pint 0.10nA)}.
\end{figure}

Figure~2(a) shows a topographic STM image in a 40~nm$\times$40~nm area. In
the measured set of $dI/dV$ images with different sample bias voltages clear
QPI features around the Co adatoms are observed. The QPI standing wave gradually
changes its wave length depending on $V_{s}$, see Figs.~2(b)-2(f). As $V_{s}$
decreases, the wavelength in the real space ($r$-space) increases. For
$V_{s}<200$~mV, the interference patterns become
rather diffuse. Moreover, the standing waves observed around the Co
adatoms exhibit strong anisotropic shape, in contrast to what is commonly observed
at high-index surfaces of pure metals \cite{Cu_SW_nature1993,Nishimura2009PRB}.

\begin{figure}
\includegraphics[width=\columnwidth]{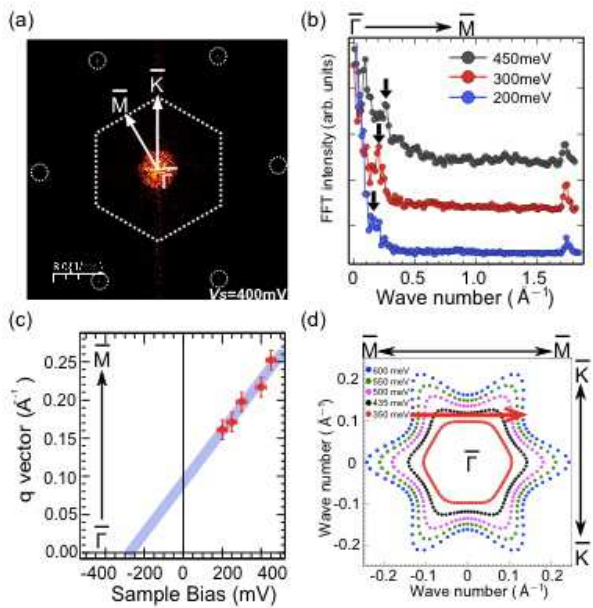}
\caption{\label{fig:epsart2} (Color online) (a) Fourier transformed $dI/dV$ images
  acquired at $V_{S}$ = 400 mV, surface Brillouin zone is indicated by
  the white dashed hexagon. (b) Profile curves of Fourier transformed
  image along $\bar{\Gamma}$-$\bar{\rm M}$ at different sample bias
  voltages.  (c) Plot of the interference peak along
  $\bar{\Gamma}$-$\bar{\rm M}$ direction (indicated by an arrow in
  (c)) as a function of sample bias voltages. (d) Calculated constant
  energy contour of Bi$_{2}$Se$_{3}$, the red arrow indicates the
  scattering channel along $ \bar{\Gamma}-\bar{\rm K}$ due to the
  perpendicular spin component.}
\end{figure}

To gain a deeper insight into the relationship between the interference
pattern and $V_{s}$, the real space $dI/dV$ maps are Fourier transformed,
as depicted in Fig.~3(a) for the sample bias voltage of 400~mV. The six
spots marked with white dashed circles are (1$\times$1) spots in the
reciprocal space, and the (0, 0) spot is located in the center of the
image. The surface Brillouin zone is shown in Fig.~3(a) as a hexagon with
the $\bar{\Gamma}$ point located in the center. The reciprocal space image
in Fig.~3(a) reveals the anisotropic interference patterns with six-fold
symmetry. The intensity maximum is in the $\bar{\Gamma}$-$\bar{\rm M}$
direction as shown by the cross-sectional profiles at different sample bias
voltages in Fig.~3(b). The six-fold pattern gradually shrinks with decreasing
$V_{s}$ and eventually vanishes below 200~mV, as indicated by arrows in
Fig.~3(b). Further quantitative analysis reveals that wave number of the
peak in the cross-sectional profiles along $\bar{\Gamma}$-$\bar{\rm M}$
almost linearly shifts with $V_{s}$ [Fig.~3(c)].

In case that the TR symmetry is broken, the appearance of the energy gap at
the Dirac point would simultaneously result in the opening of the scattering
channels at all energies.
Thus, the vanishing of the standing waves below 200~mV in the $dI/dV$ map, where the warping effect of Dirac cone becomes neglectable, 
suggests that the TR protection of the topological
surface state persists without a gap opening at the Dirac point even in the
presence of Co impurities. The scattering observed with large sample bias
voltages in the unoccupied states can be ascribed to a heavily deformed
iso-energy contour in the high energy range (the unoccupied states) as indicated
by the scattering vector superimposed on the calculated constant energy contours
in Fig.~3(d). The scattering channel is allowed because of the finite surface
perpendicular spin component \cite{kuroda2010BiSe,Fu2009PRL}.

To see how the surface states dispersion is modified upon Co deposition, we
have employed ARPES with
synchrotron radiation. Two sets of measurements were performed with different
Co deposition durations on two samples labeled A and B. Here, we use the
photon energy of 50~eV, at which the spectral weight of the bulk conduction
band is largely suppressed and the surface Dirac cone is highlighted.

\begin{figure}
\includegraphics[width=\columnwidth]{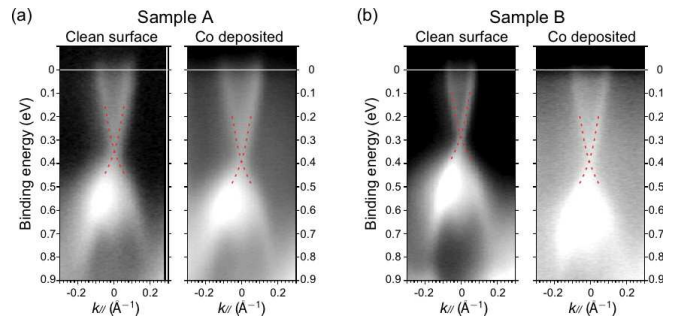}
\caption{\label{fig:epsart3} (Color online) ARPES measurement on two samples
before and after (a) 0.2 ML and (b) 0.9 ML
of Co deposited acquired by photon energy of 50 eV. }
\end{figure}

For both samples, the linear surface state dispersion is clearly observed
before as well as after the Co deposition, but the background is
stronger with larger Co coverage. For the sample A [Fig.~4(a)], the Dirac
point prior to the Co deposition is located at 340~meV below the Fermi level.
After 0.2~ML Co deposition, the Dirac point shifts to 390~meV, and no other
significant changes of the Dirac cone is observed. When the amount of Co
increases to 0.9~ML (sample B, Fig.~4(b)), a larger energy shift
of $\sim 100$~meV is observed with stronger background intensity. Again,
neither the change of spectral feature nor the energy gap opening is observed.
These results imply either the absence of the ferromagnetic order or the
vanishing of the magnetic moment of Co atoms deposited on the surface
of Bi$_{2}$Se$_{3}$, which may be ascribed to the chemical bonding with Se atoms
on the topmost layer. Thus, the QPI observed in the present STM results should be originated from the deformation of the Fermi surface
\cite{kuroda2010BiSe, Fu2009PRL} rather than from a broken TR symmetry.

\begin{figure}
\includegraphics[width=\columnwidth]{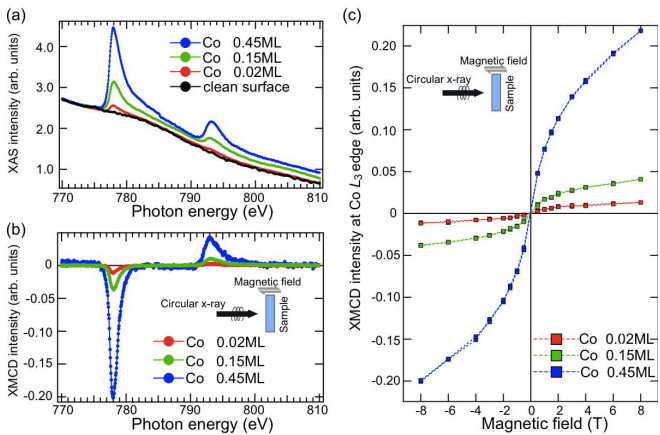}
\caption{\label{fig:epsart4} (Color online) (a) XAS, (b) XMCD spectra at Co $L_{23}$
  edges, and (c) site-specific magnetization measurement of Co $L_{3}$
  edge with different coverage of Co on Bi$_{2}$Se$_{3}$ surface
  (temperature, 5 K).  }
\end{figure}

Finally, to understand the magnetic properties of Co on the surface of
Bi$_{2}$Se$_{3}$, we have performed X-ray magnetic circular dichroism
measurement, which is able to resolve element-specific magnetic properties.
Figure~5(a) shows the Co $L_{23}$ X-ray absorption spectra (XAS) for three
different coverages. Also the absorption spectrum of the clean sample is measured to estimate the background contribution.
As can be seen in Fig.~5(a), when the coverage of Co increases from 0.02 to
0.45~ML, the total absorption intensities at 778 and 793~eV gradually increase.
The XMCD spectrum is obtained by measuring the difference of the absorption
intensities with the circular polarization vector of the incident light
parallel and anti-parallel to the magnetic field. Figure~5(b) shows the
XMCD spectra acquired under 8~Tesla magnetic field applied perpendicular
to the sample surface at 5~K. The intensity of XMCD signal of both $L_3$
and $L_2$ edges increases with increasing Co coverage. Note that
the ratio between the orbital and spin magnetic moment deduced from the
XMCD spectra of the Co layer deposited on Bi$_{2}$Se$_{3}$
($m_{\rm orb}/m_{\rm spin}\sim 0.3$--0.5) is for the three measured Co
coverages considerably larger than for the bulk Co crystal (
$m_{\rm orb}/m_{\rm spin}\sim 0.1$) \cite{Chen1997PRL}.

Whether the Co deposited Bi$_{2}$Se$_{3}$ surfaces are ferromagnetic or not
can be inferred by investigating site-specific magnetization as a function
of magnetic field. Figure~5(c) shows the intensity of XMCD signal at Co $L_3$
edges as a function of the magnetic field (M-H curve), which is swept from
$-8$ to 8~T and then back to $-8$~T to form a loop. As is seen from Fig.~5(c),
the M-H curves for all the three coverages of Co increase in the same
direction with increasing the magnetic field strength. However, the intensity
of the M-H curves are not saturated even at 8~T and no magnetic hysteresis is
observed for any of the three Co coverages. These results clearly show that
the Co layer on the Bi$_{2}$Se$_{3}$ surface has no long-range ferromagnetic order. This is consistent with our ARPES results, which show
no gap-opening at the Dirac point upon Co deposition, and, thereby, confirm
that the QPI patterns observed in the STM experiment originate from the
strongly warped Dirac cone in the unoccupied states of Bi$_{2}$Se$_{3}$.

\begin{figure}
\includegraphics[width=0.7\columnwidth]{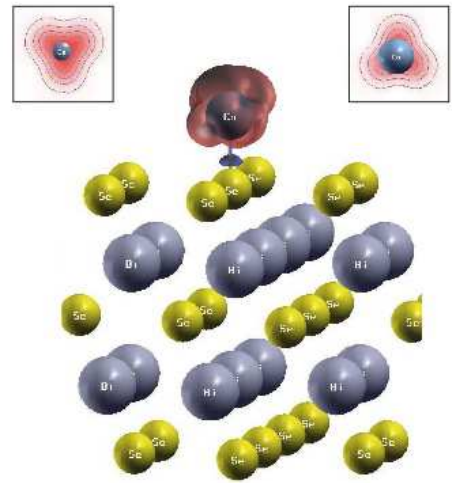}
\caption{(Color online) Constant-density surface for perpendicular ($z$)
magnetization density (red and blue colors correspond to positive
and negative values, respectively). Insets show cross-sections of
this surface in $xy$ plane just above (left) and below (right) the
Co atom center.}
 \label{fig6}
\end{figure}

In order to further investigate the magnetic state of cobalt on the
surface, we have performed {\em ab initio} DFT calculations using the
VASP code \cite{VASP1,VASP2,PAW2}. We considered a single Co adatom
within a slab model of a $3\times 3$ surface super-cell of
Bi$_2$Se$_3$. The lateral position of the atom was fixed to be above
the Se atom in accordance with our STM data, and the height of the
Co atom above the top Se atom was optimized to be 2.27~\AA\ (within
a collinear scalar-relativistic approximation). The magnetic
properties of the Co adatom in this position were calculated with
the inclusion of the spin-orbit coupling. The magnetic moment of the
Co adatom is found to be 2.06~$\mu_{\rm B}$ along the $z$ axis. At
the same time, the magnetic adatom does not induce a noticeable spin
polarization of the substrate. The calculated $z$-magnetization
density is shown in Fig.~\ref{fig6}. The obtained negligible
polarization of the substrate explains why the  magnetic interaction
between the adsorbed Co atoms is extremely weak, and why no
ferromagnetic ordering at a low Co coverage is observed in the
experiment.

In summary, by measuring the bias-dependent differential conductance mapping
we have observed the QPI of Dirac electrons induced by Co impurities in the unoccupied states of Bi$_{2}$Se$_{3}$. The
magnetic properties of Co deposited Bi$_{2}$Se$_{3}$ surfaces investigated
by XMCD reveal an absence of ferromagnetism, in agreement with our ARPES
results that the Dirac point of the topological surface state is not destroyed
by the Co impurities. We conclude that the QPI near Co impurities originates
from the strongly deformed isoenergy contours of the unoccupied states of
Bi$_{2}$Se$_{3}$. The present results may also provide a pathway to future studies
of Kondo physics in the presence of the topological surface electrons.

\begin{acknowledgments}
STM and ARPES measurements were performed with the approval of the Proposal Assessing Committee of HSRC (Proposal No.11-B-40, No.10-A-32). The XMCD experiment was performed at SPring-8 with the approval of Japan Atomic Energy Agency (JAEA) as Nanotechnology Network Project of the Ministry of Education, Culture, Sports, Science and Technology. (Proposal No. 2011A3873/BL23SU). This work was partly supported by KAKENHI (No. 20340092, 23340105), Grant-in-Aid for Scientific Research (B) of Japan Society for the Promotion of Science.  Author MY thanks financial support from JSPS Research Fellowship.
\end{acknowledgments}

\bibliography{apssamp}

\end{document}